\documentclass[11pt,a4paper,american]{article}
\usepackage{jinstpub}
\usepackage[T1]{fontenc}
\usepackage[utf8]{inputenc}
\usepackage{babel}
\usepackage{float}
\usepackage{textcomp}
\usepackage{amsmath}
\usepackage{amssymb}
\usepackage{upgreek}
\usepackage{graphicx}
\usepackage{gensymb}
\usepackage{eurosym}
\usepackage{tikz}
\usepackage{subcaption}
\usepackage[super]{nth}
\usepackage{todonotes}
\usepackage{physics}
\usepackage[per-mode=symbol,separate-uncertainty=true,exponent-product=\cdot,multi-part-units=brackets,product-units=power]{siunitx}
\DeclareSIUnit\depth{\gram\per\square\centi\meter}
\usepackage{afterpage}
\usepackage[section,above,below]{placeins}
\usepackage{xspace}
\usepackage{hyperref}
\usepackage{tikz}

\renewcommand{\[}{\begin{equation}}
\renewcommand{\]}{\end{equation}}

\newcommand{\image}[1]{\node[anchor=south west,inner sep=0] (image) at (0,0){#1};}
\newcommand{\entry}[1]{\raisebox{2pt}{\tikz{\draw[line width=2pt,#1] (0,0) -- (5mm,0);}}}
\newcommand{\header}[1]{\multicolumn{2}{l}{\raisebox{1.5pt}{#1}}}
\newcommand{\legend}[2]{ 
\node[draw=black,fill=white,anchor=south west](legend) at (#1) 
{\begin{tabular}{@{}r@{ }l@{}}#2\end{tabular}};}
\title{Zenith angle dependence of the cosmic ray rate as measured with Imaging air-Cherenkov Telescopes}
\author[a]{Thomas Bretz}
\affiliation[a]{III. Physikalisches Institut, RWTH Aachen University \\ Otto-Blumenthal-Straße, 52074 Aachen, Germany}
\emailAdd{tbretz@physik.rwth-aachen.de}

\abstract
{The rate of extensive air-showers observed with imaging air-Cherenkov telescopes is zenith angle dependent. This effect originates from the increasing geometrical distance of the observed shower to the telescope with increasing zenith distance. This paper investigates how this alters the observed image and how this affects the trigger rate as a function of zenith angle. The discussed effects include the change of Cherenkov light yield, of absorption in the atmosphere, of photon density at the aperture and of the image size at the focal plane of the telescope. Based on a simple model for the atmosphere and well-known first principles on the development of extensive air-showers, the zenith angle dependence is expressed analytically. The assumption that most light is emitted from the shower core and mathematical approximations allow to derive an analytical expression describing the zenith angle dependence well with only three free parameters which are directly linked with the underlying physics. This suggests further investigations about how these fit parameters are linked to the properties of the atmosphere and the instrument. Using data published by the First G-APD Cherenkov Telescope, a good match of the fit functions with the data is obtained. For the trigger rate of cosmic rays, the obtained parameters are consistent with the naive expectation.}

\makeatletter
\let\thetitle\@title
\let\themail\@mail
\let\theauthor\@author
\makeatother

\begin{document}
\maketitle

%\newpage

%\listoftodos

%\newpage

%https://www.fact-project.org/dch/db_explorer.php?short=3628
%https://www.fact-project.org/dch/db_explorer.php?short=3614

%%%%%%%%%%%%%%%%%%%%%%%%%%%%%%%%%%%%%%%%%%%%%%%%%%%%%%%%%%%%%%%%%%%%%%%%%%%%%%%%%%%%
%%%%%%%%%%%%%%%%%%%%%%%%%%%%%%%%%%%%%%%%%%%%%%%%%%%%%%%%%%%%%%%%%%%%%%%%%%%%%%%%%%%%
\section{Introduction}

% Introduction to imaging air-Cherenkov telescopes in general
% Introduction to the physics

During the past decades, observations at TeV energies with imaging air-Cherenkov telescopes became an important part of current astronomy and provide valuable data for the understanding of all kind of sources and eventually the origin of very high- and ultra-high energy cosmic rays~\cite{bib:review}.

Imaging air-Cherenkov telescopes observe Cherenkov light emitted by particle cascades in the atmosphere. These cascades are induced by primary charged particles and gamma-rays.

A common problem for the operation of imaging air-Cherenkov telescopes is the assessment of their data quality. Data quality depends strongly on the atmosphere and thus on many variables like weather condition, but also on the performance of the optical system and the applied photon detectors. Monitoring of the telescope performance and the atmosphere in detail can be quite expensive and time-consuming.

Moreover, Cherenkov telescopes observe the spectrum of cosmic rays as a natural background above several GeV and consequently beyond solar modulation effects, At these energies, the spectrum is known as one of the most stable fluxes ever measured~\cite{bib:pdl}. Consequently, every performance change in the detector or change of atmospheric condition directly alters the measured trigger rate through the change in detected light yield. %Consequently, a measured change Because the change of the flux of charged cosmic rays is very well known, a change in rate can also be converted to a change in energy which is directly related to a change in the measured light yield.

Apart from systematic changes in the detection efficiency, the measured rate of the system depends on the hardware threshold of the trigger system and on the zenith angle at which observations take place. If the hardware threshold is adapted to background light conditions, as it is done for the FACT telescope to achieve the lowest possible energy threshold, the trigger rate is altered directly~\cite[e.g.][]{bib:current}. Observations carried out at different zenith angles observe showers of the same energy developing inclined in the atmosphere at increased geometrical distance. Therefore, observing the same spectrum at identical atmospheric and hardware conditions indirectly alters the trigger rate. This effect, of course, is predictable and has a well defined dependency on the observation angle.

If adaption of the hardware trigger threshold takes place, assessing the measured trigger rate directly is difficult. Instead, a software trigger can be implemented with a trigger threshold well above the hardware threshold of all observations. The measured rate of events surviving this software trigger is then independent of the hardware trigger threshold and can be understood as a direct measurement of the cosmic ray rate. Such a software trigger was first proposed in~\cite{bib:hildebrand1} based on~\cite{bib:hildebrand2} to mimic the hardware trigger with an adjustable trigger threshold. Data at a fixed zenith angle can then be used to find a threshold for which the measured rate is independent of the applied hardware threshold settings. This provides a well-defined dependence of the measured rate on zenith angle and on light yield. 

The development of extensive air-showers is independent of the penetrated medium and depends only on the column density. Consequently, air-showers can be described by an energy dependent profile if their geometry is expressed in units of the integrated density, the so-called {\em atmospheric depth}. For inclined showers, this results in a shower development higher up in the atmosphere and an adapted geometrical development. As a consequence, the light yield and illuminated area on the ground also changes. These effects lead to the well-known change of the measured rate with zenith angle and are discussed in more details in Sec.~\ref{sec:model}.

Once, the zenith angle dependence is known, measured rates can be compared with the rates measured at optimal conditions and a single quality parameter can be derived. Although, today's understanding of extended air-showers is well advanced, their underlying physics processes are too complicated to be assessed by a precise mathematical model and Monte Carlo simulation are required. Simplified analytical expressions derived from first principles can describe the average shower behavior well enough to deduce mathematical models which fit the data reasonably well. These simplifications often require adapting of constants but still lead to useful results.\\

In previous publications, a phenomenological approach was used to link the measured trigger rate with properties of the atmosphere~\cite[e.g.][]{bib:stefanik,bib:hahn}. The goal of the presented study is to find a physics driven fit model to describe the measured trigger rates for hadron induced showers (background) and gamma induced events (signal) as a function of the observed zenith angle $\theta$. A first approach has been carried out in~\cite{bib:mahlke} based on simple shower physics. This paper will comprehensively extend the model to derive at a precise description of the data without systematic deviations. In analogy to the cited papers, the obtained fit parameters carry valuable information on the performance of the instrument and the properties of the atmosphere. However, their interpretation is beyond the scope of this paper.\\

In the following, all important physics and geometrical effects are discussed and combined into a single fit formula with three parameters. All effects are discussed independently of each other. To derive an analytical fit formula, approximations are performed which are separated from the pure analytical description by dedicated paragraphs.

%%%%%%%%%%%%%%%%%%%%%%%%%%%%%%%%%%%%%%%%%%%%%%%%%%

\section{The physics model}\label{sec:model}

\paragraph{Definitions} The zenith angle $\theta$ describes the inclination of the optical axis (also `line-of-sight') of a telescope with respect to zenith. The collection area of the optical system (e.~g.\ reflector or lens) is called {\em aperture}. The plane spanned by the aperture perpendicular to the line-of-sight is called {\em aperture plane} in the following. The instrument in the focal plane of the telescope optics detecting photons from the shower is called {\em camera}.%\todo{Is there a better word?}

\paragraph{Trigger Rate} The trigger rate of a telescope is directly linked to the primary particle spectrum~$\phi(E, \Theta)$. In a first approximation, the trigger rate $R(\theta)$ can be obtained by the integral starting at an effective energy threshold $E_\text{th}$ to infinity over the differential flux $\phi(E, \theta)$ times the effective collection area $A_\text{eff}(E, \theta)$ and the observed solid angle $\Omega(E, \theta)$:
\[\label{eq:rate}
R(\theta) = \int\limits_{E_\text{th}(\theta)}^\infty \phi(E, \theta)\cdot A_\text{eff}(E, \theta) \,\dd{E} \]

For cosmic rays, within the required precision, the flux is usually homogeneous and can be described as a power law with spectral index $\gamma$ and thus
\[\label{eq:spectrum}
\phi(E, \theta) \equiv \phi(E) \propto E^{-\gamma}\]

The effective collection area $A(E, \theta)$ corresponds to the area illuminated on the aperture plane by a shower. It depends on the energy $E$ of the primary because the optical depth of the shower development in the atmosphere changes and therefore its distance to the aperture plane. It also depends on the inclination angle at which a shower hits the Earth's atmosphere, because the same optical depth would be reached at higher altitudes above ground for more inclined showers. 

%In the following, a flat Earth approximation is made. Knowing that this approximation is valid up to at least $\theta$=75\textdegree{}, the inclination angle of a shower parallel to the line-of-sight with the atmosphere is equal to the zenith angle $\theta$. In almost all cases, Cherenkov telescopes observe well below $\theta$=65\textdegree{} where this approximation is fully valid precisely due to the strong decrease in sensitivity. Therefore, the goal of this study is mainly to describe the trigger-rate up to about 70\textdegree{}. More detailed studies might be required for an analytical description beyond that.

The solid angle $\Omega(E, \theta)$ corresponds to all directions around the line-of-sight from which showers are visible. A shift of the shower axis parallel to the line-of-sight mainly moves the shower out of the field-of-view of the camera.%, and shower inclined by more than an angle $\delta$ with respect to an axis parallel to the line-of-sight larger than the angle $\vartheta_\text{C}$ at which the Cherenkov light is emitted by the shower will not be visible anymore.\todo{Light is the center comes from the top of the shower}

To estimate this change of energy threshold with zenith angle, the effective collection area and the geometry of the shower development in the atmosphere needs to be described.

\subsection{Atmospheric Model}\label{sec:atmosphere}%\todo{Maybe there is a publication with an analytical model?}

As shown later, the basic properties of the atmosphere required, are its temperature, its density and its pressure profile as a function of height above ground.

% From: https://usatoday30.usatoday.com/weather/wstdatmo.htm
For an accurate description of the temperate $T(h)$, pressure $p(h)$ and density $\rho(h)$ as a function of height $h$ above ground, a hypothetical vertical distribution of atmospheric properties which, by international agreement, is roughly representative of year-round, mid-latitude conditions has been used. The so-called {\em U.~S.~Standard Atmosphere 1976}~\cite{bib:atmosphere} is an idealized, steady state representation of the Earth's atmosphere during a period of moderate solar activity. 

To be able to apply them more easily in numerical calculations, all three properties were fitted with some arbitrary polynomial and exponential functions which will be used in the following:
\[\label{eq:temperature}
\frac{T(\frac{h}{\text{km}})}{\degree\text{C}}
= 14.88 - 5.11\cdot h - 0.649\cdot h^2 + 0.0767\cdot h^3 - 0.00261\cdot h^4 + 0.000294\cdot h^5
\]
\[\label{eq:density}
\frac{\rho(\frac{h}{\text{km}})}{ \frac{\text{kg}}{\text{m}^3}}
= 1.17881 \cdot\text{e}^{-0.00642\cdot h^{1.22649}} 
\]
\[\label{eq:pressure}
\frac{p(\frac{h}{\text{km}})}{\text{hPa}}
= 1007\cdot\text{e}^{-0.106\cdot h^{1.1043}}
\]

\begin{figure}
\centering
\includegraphics[width=.49\textwidth]{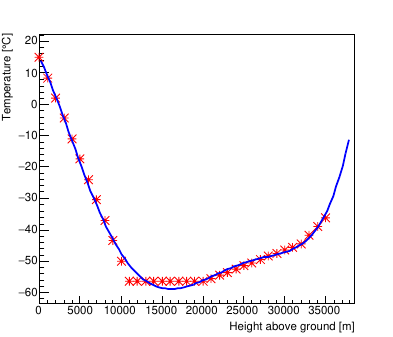}
\includegraphics[width=.49\textwidth]{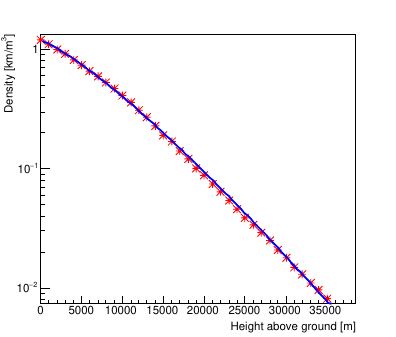}
\includegraphics[width=.49\textwidth]{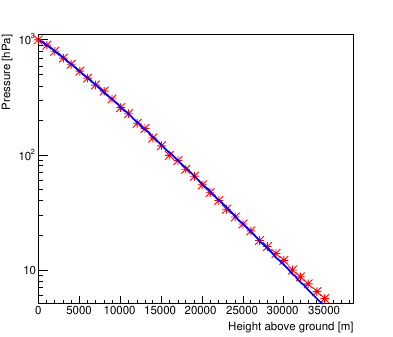}
\caption{\label{fig:Tdp-H}Temperature, density and pressure as a function of height as defined in the U.~S.~Standard Atmosphere 1976. The red lines are fits to the data as discussed in the text.}
\end{figure}

The U.~.S. Standard Atmosphere and the fits are shown in Fig.~\ref{fig:Tdp-H}.

%\afterpage{\clearpage}

\subsection{Refractive index}\label{sec:refractive-index}

From the atmosphere's pressure and temperature profile, the refractive index for a given wavelength~$\lambda$ can be approximated precisely in the wavelength range between \SI{405}{\nano\meter} and \SI{705}{\nano\meter}~\cite[c.~f.][]{bib:birch} as
%
% From: http://www.kayelaby.npl.co.uk/general_physics/2_5/2_5_7.html
\[\label{eq:refractive-index}
n(\frac{p}{\text{Pa}}, \frac{T}{\degree\text{C}}) \approx 1 + n^* \cdot f(p, T)\]

\noindent with
\[
f(p, T)\equiv
\frac{1+p\cdot(60.1-0.972\cdot T)\cdot10^{-10}}{96\,095.43\cdot(1+0.003661\cdot T)}
\]

\noindent and
\[n^* \approx \frac{0.0472326}{173.3-\frac{1}{\lambda^2}}\]

\begin{figure}
\centering
\begin{tikzpicture}
\sffamily
\image{\includegraphics[width=.68\textwidth]{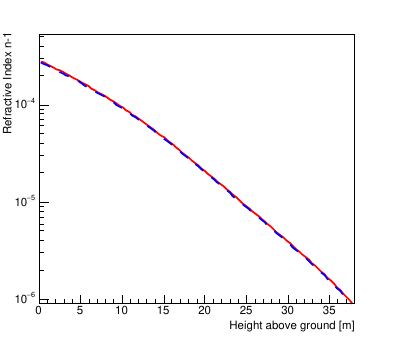}}
\legend{4.4,6}{
  \header{U.S. Standard Atmosphere}\\
  \entry{dashed,black}&\SI{400}{\nano\meter}\\
  \entry{solid,red}&\SI{700}{\nano\meter}\\
}
\end{tikzpicture}
\caption{\label{fig:refractive-index}Refractive index as a function of height above ground for the U.~S.~Standard atmosphere for a wavelength \SI{400}{\nano\meter} (solid, red) and \SI{700}{\nano\meter} (dashed, blue).}
\end{figure}

Although, the sensitive wavelength range of a Cherenkov telescope lays partially outside of this range starting at around \SI{280}{\nano\meter} defined by the UV-cutoff in the atmosphere and extending up to \SI{900}{\nano\meter} for modern photo sensors, for this study, the loss of precision outside of the formula's validity range is still negligible as the change with height dominates. Fig.~\ref{fig:refractive-index} shows the refractive index as a function of height above ground as derived from Eq.~\ref{eq:refractive-index} and from the equations from Sec.~\ref{sec:atmosphere}.
    
%\afterpage{\clearpage}

\subsection{Atmospheric depth}\label{sec:atmospheric-depth}

The mean interaction lengths of photons and particles penetrating a medium does not depend on geometrical distance but on the amount of matter penetrated. Therefore, the atmospheric depth has been defined as a measure for the mass per surface area along the line-of-sight.

The slant atmospheric depth $X(d)$ at distance $d$ along the line-of-sight is given by the integral over the atmospheric density $\rho$ which is dependent on height above ground $h$
\[\label{eq:slant-depth}
X(d) = \int\limits_d^\infty\rho(h(d))\cdot\dd d\]

Under an inclination angle $\theta$, height above ground becomes a function of slant distance $d$
\[\label{eq:curved}
h(d) = d\cdot\cos\theta + \frac{d^2\sin^2\theta}{2R}
\]

where $R$ is the Earth's radius. The atmospheric depth $X(d)$ becomes $X(d, \theta)$. It is apparent that for a line-of-sight perpendicular to ground ($\theta$=0\textdegree{}), $h(d)$ is equivalent to $d$. These equations will be referred to as the {\em curved atmosphere} in the following.%\todo{Reference}

Approximating the Earth as flat, $h(d)$ becomes $d\cdot\cos\theta$, and the atmospheric depth can be simplified to
\[X(d, \theta) \approx \frac{X(h(d))}{\cos\theta}\]

%This approximation is usually valid up to $\sim$80\textdegree{}.

Assuming an ideal gas and a constant temperature, Eq.~\ref{eq:slant-depth} can be solved. This leads to a purely exponential pressure profile of the atmosphere which is called {\em barometric} and {\em flat} in the following.

The assumption of a flat Earth then yields the following expression
\[\label{eq:height-flat}
h(X, \theta) \approx -h_0\cdot \ln\frac{X\cdot\cos\theta}{X_0}
\]

Here, $h_0$ is the scale height of the atmosphere of \SI{7.7}{\kilo\meter} and $X_0$\,=\,\SI{1035}{\depth} the total depth of the vertical atmosphere. This equation will be referred to as the {\em flat atmosphere} in the following. Note that in this context, the flat atmosphere is not just a geometrical term but already includes the barometric profile.

\begin{figure}
\centering
\begin{tikzpicture}
\footnotesize\sffamily
\image{\includegraphics[width=.495\textwidth]{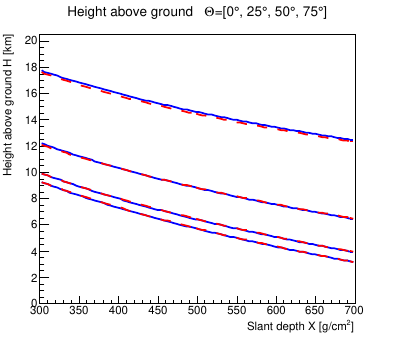}}
\legend{4.5,4.5}{
  \entry{dashed,red}&\small Numeric\\
  \entry{solid,blue}&\small Fitted ($*$)\\
}
\end{tikzpicture}
\begin{tikzpicture}
\footnotesize\sffamily
\image{\includegraphics[width=.495\textwidth]{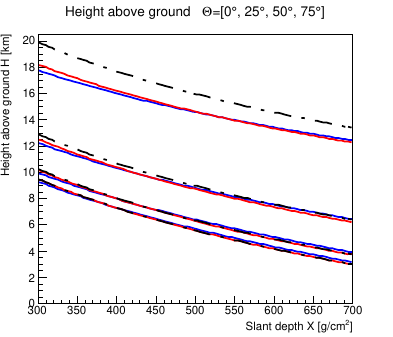}}
%\legend{0.9,0.8}{
\legend{4.7,4.5}{
  \entry{dashdotted,black}&Flat,\SI{350}{\depth}\\
  \entry{solid,blue}&Curved ($*$)\\
  \entry{solid,red}&Adjusted\\
}
\end{tikzpicture}
\begin{tikzpicture}
\footnotesize\sffamily
\image{\includegraphics[width=.495\textwidth]{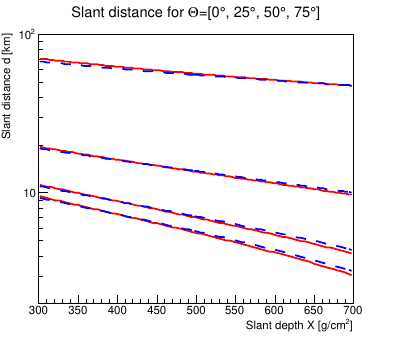}}
\legend{4.6,3.4}{
  \entry{dashed,blue}&Curved\\
  \entry{solid,red}&Adjusted\\
}
\end{tikzpicture}
\begin{tikzpicture}
\footnotesize\sffamily
\image{\includegraphics[width=.495\textwidth]{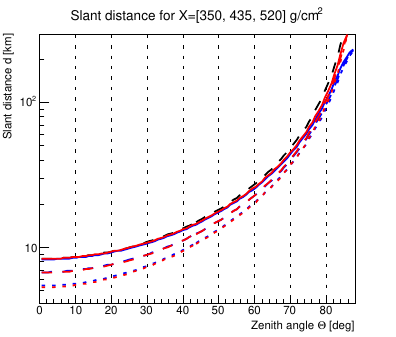}}
\legend{1.0, 4.0}{
  \entry{dashdotted,black}&Flat, \SI{350}{\depth}\\
  \entry{solid,blue}&Curved\\
  \entry{solid,red}&Adjusted\\
}
\end{tikzpicture}
\caption{\label{fig:H-X}
{\bf Top, Left:} A comparison of between the {\em U.S. Standard Atmosphere} (red) with the fit-model (blue). Both are calculated using a curved atmosphere as described in Eq.~\ref{eq:curved}. They are in very good agreement.
{\bf Top, Right:} A comparison between the curved atmosphere (blue) using the fit model, the flat barometric atmosphere (black, dashed) and the adjusted atmosphere~(red). While significant deviations of the flat barometric from  the curved atmosphere are visible, the adjusted atmosphere is in good agreement.
{\bf Bottom, Left:} Slant distance as a function of slant atmospheric depth for different inclination angles between \SI{0}{\degree} (bottom) and \SI{75}{\degree} (top). A good agreement for the curved atmosphere (blue, dashed) and the adjusted atmosphere (red) is visible.
{\bf Bottom, Right:} Slant distance as a function of inclination angle for different atmospheric depth of \SI{350}{\depth} (solid), \SI{435}{\depth} (dashed) and \SI{520}{\depth} (dotted) calculated for the curved (blue) and the adjusted atmosphere (red). They are in good agreement up to at least \SI{70}{\degree}. For reference, also the flat atmosphere is shown (black, dashed) for \SI{350}{\depth}.}
\end{figure}

Phenomenologically, it can be found that an adjusted equation
\[\label{eq:height-adjusted}
h(X, \theta) \approx -h_0\,(\cos\theta)^\alpha\cdot \ln\frac{X\cdot\cos\theta}{X_0}
\]

with an arbitrary exponent $\alpha=0.065$ gives a better description than Eq.~\ref{eq:height-flat}, see Fig.~\ref{fig:H-X}. In the following this equation will be referred to as the {\em adjusted atmosphere}.

The flat and adjusted atmosphere have the advantage of fast numerical calculations while the curved atmosphere takes significantly more computing time.

A comparison of all the different descriptions of the atmosphere can be found in Fig.~\ref{fig:H-X}. It can be seen that the adjusted equation gives a much better match with the curved U.S. Standard atmosphere above an inclination angle of \SI{50}{\degree} compared to the pure flat barometric model and yields reasonable results up to at least \SI{70}{\degree}. Therefore, in the following the adjusted atmosphere will be used.

In most cases, quantities calculated in the following will be shown for the adjusted and the flat atmosphere to show that they provide almost identical results and deviations start not earlier than \SI{70}{\degree}. For comparison, examples of the flat atmosphere are shown. They usually start to deviate around \SI{45}{\degree} and become significant already above \SI{60}{\degree}.

%\afterpage{\clearpage}

%The distance from the telescope to a given optical depth $X$ as seen under a given zenith angle $\theta$ is thus

%\[\label{eq:distance}
%D(\theta) = \frac{H(\theta)}{\cos\theta}
%\]

%The distance to the shower core calculated using the exact equation Eq.~\ref{eq:slant-depth} is shown in Fig.~\ref{fig:slant-depth} (red). The approximation from Eq.~\ref{eq:distance} is shown in blue. They agree well up to about 80\textdegree{}. This agreement can further be improved including an additional exponent

%\[\label{eq:distance-fit}
%D(\theta) = \frac{H(\theta)}{\cos\theta^{0.93}}
%\]

%\begin{figure}
%\centering
%\end{figure}
%\afterpage{\clearpage}

\subsection{Shower maximum}\label{sec:shower-maximum}

Most of the light is emitted at the shower maximum at the atmospheric depth $X_\text{max}$. The atmospheric depth of the shower maximum is dependent on the energy $E$ of the primary particle. 

According to the Heitler-Model~\cite{bib:heitler}, the atmospheric depth for the maximum of electromagnetic showers can be estimated using a simple cascade model as
%\[X^\gamma_\text{max}(E)=\lambda_\text{e}\cdot\left(\frac{1}{2}+\frac{\ln\frac{E}{E_\text{c}}}{\ln2}\right) \label{eq:heitler}\]
\[X^\gamma_\text{max}(E)\approx\lambda_\text{e}\cdot\left(\frac{1}{2}+\ln\frac{E}{E_\text{c}}\right) \label{eq:heitler}\]

where $E_\text{c}$ is the critical energy in air of 86\,MeV and $\lambda_\text{e}$ is the typical interaction length of the electrons and photons in the shower of about \SI{37.4}{\depth}. The additional factor \textonehalf{} compared with the cited solution arises from a more detailed computation considering the energy distribution of particles in the shower~\cite[][Eq.~4.14]{bib:spurio}.

% Astroparticle physics 22 (2005)  387-297
% http://particle.astro.ru.nl/ps/astropart1415-wk7a.pdf
Matthews~\cite[][Eq.~12]{bib:matthews} extended the model for proton induced showers. Just for an easier comparison with Eq.~\ref{eq:heitler}, his solution was rephrased as
\[X^\text{p}_{max}(E) \approx \lambda_\text{p}\cdot\left(6.9+\ln\frac{E}{E_\text{c}}\right)\] %% log_10 !!!!

\noindent with $\lambda_{p}\equiv$\,\SI{25.1}{\depth}.\\ %The value for $X_\text{s}$ was already corrected for the effects discussed in the text.

% Xgammamax = nc lambdar ln(2)
% Xpmax = Xgammamax + X0 - lambdar*ln(3Nch)
% Eq. 12: xpmax = X0 + lambdar ln (E0 / 3Nch zeta)
% X0 ~ 61g/cm^2 * ln(2)
% Eq 12: 470 + 58 log10(E/1PeV)
% 

\begin{figure}
\centering
\begin{tikzpicture}
\sffamily
\image{\includegraphics[width=.68\textwidth]{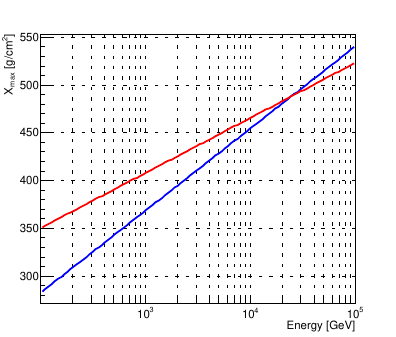}}
\legend{1.7, 6.}{
  \entry{solid,blue}&Gammas\\
  \entry{solid,red}&Protons\\
}
\end{tikzpicture}
\caption{\label{fig:xmax-E}Atmospheric depth of the shower maximum for gammas (blue) and protons (red) as a function of their energy according to the Heitler-/Matthews-model.}
\end{figure}

Both functions are plotted in Fig.~\ref{fig:xmax-E}. Combined with Fig.~\ref{fig:H-X}, this also shows that for the whole energy range of a typical Cherenkov telescope from about \SI{100}{GeV} to about \SI{100}{TeV}, the shower maximum is well above \SI{2}{\kilo\meter} and does not reach the aperture plane.%\todo{comment later on the energy of TH750 -- use Rate as reference}

According to the Heitler-Model and simulations usually carried out for the analysis of the data from imaging air-Cherenkov telescopes, the number of particles in a shower is proportional to the primary energy. Consequently, as the emitted Cherenkov light yield does not depend on the particle's energy, the number of emitted Cherenkov photons is also directly proportional to the primary energy. Thus, in the following, the Cherenkov light yield and the energy of the primary particle can be considered interchangeably. For hadronic primaries, this is only true in a first approximation, but can be considered valid over the comparably small energy range considered here.

%\afterpage{\clearpage}

%\section{Shower height}\label{sec:shower-height}

%For vertical showers, the geometrical height $H_\text{max}$ of the shower core above ground is obtained from its atmospheric depth $X_\text{max}$ as

%\[\label{eq:hmax}
%H_\text{max}(E, \theta) = -H_0\cdot \ln\frac{X_\text{max}(E)}{X_\text{tot}(\theta)}\].

%assuming a barometric atmosphere.\todo{Include a correct profile using the approximation above} Then, $H_0$ is the scale height of the atmosphere. It is about 6.5\,km up to 10\,km to 15\,km and about 9.5\,km above. The term $X_\text{tot}(\theta)$ describes the total depth of the atmosphere along the line-of-sight

%\[\label{eq:xtot}
%X_\text{tot}(\theta) = \frac{X_0}{\cos\theta}\]

%with a total depth at $\theta=0\degree$ of $X_0$\,=\,1035\,g/cm$^2$. This equation only holds in a flat Earth approximation.

%\begin{figure}
%\centering
%\includegraphics[width=.68\textwidth]{img/plot-Hmax-Theta.png}
%\caption{\label{fig:hmax-theta}Shower height above ground as a function of zenith angle.}
%\end{figure}
%\afterpage{\clearpage}

%\section{Shower distance}\label{sec:shower-distance}

%The distance $D_\text{max}$ of the shower maximum from the telescope plane is thus

%\[\label{eq:dmax}
%D_\text{max}(E, \theta) = \frac{H_\text{max}(E, \theta)}{\cos\theta}\]

%\section{Zenith angle relation}

\subsection{Cherenkov light}\label{sec:cherenkov-light}%\todo{Include change of Cherenkov emission!}

According to the Frank-Tamm formula~\cite[e.~g.][]{bib:jackson}, the number of Cherenkov photons $N$ emitted by a particle with charge $Z$ and in a medium of refractive index $n$ per path length $x$ and wavelength $\lambda$ is proportional to
\[\frac{\dd N}{\dd \lambda\,\dd x} \propto \frac{Z^2}{\lambda^2}\left(1-\frac{1}{n^2\,\beta^2}\right)\]

\vspace{1em}Cherenkov light in an air shower is mainly emitted by electrons and positrons at relativistic energies, thus
\[\beta\approx 1\]

The steeply falling Cherenkov spectrum is cut off towards smaller wavelength by the atmospheric absorption in the ultra-violet. As shown earlier, the wavelength dependence of the refractive index is negligible compared to the density change of the atmosphere. For this paper, the spectrum is therefore assumed to be mono-energetic with a wavelength of
\[\lambda=400\,\text{nm}\]

The angle $\vartheta_{C}$ at which Cherenkov light is emitted is given by
\[\cos\vartheta_{C} = \frac{1}{n\,\beta}\approx\frac{1}{n} \label{eq:cangle}\]

\begin{figure}
\centering
\begin{tikzpicture}
\footnotesize\sffamily
\image{\includegraphics[width=.495\textwidth]{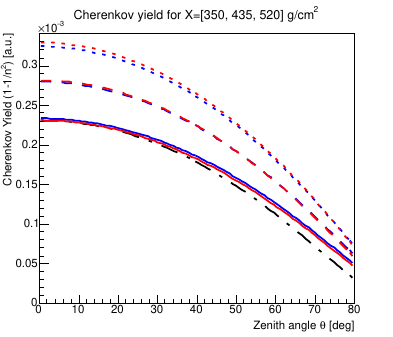}}
\legend{1.2, 1.1}{
  \entry{dashdotted,black}&Flat, \SI{350}{\depth}\\
  \entry{solid,blue}&Curved\\
  \entry{solid,red}&Adjusted\\
}
\end{tikzpicture}
\begin{tikzpicture}
\footnotesize\sffamily
\image{\includegraphics[width=.495\textwidth]{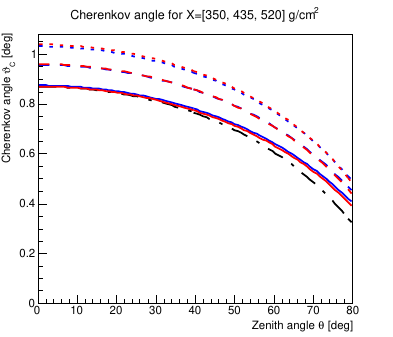}}
\legend{1.2, 1.1}{
  \entry{dashdotted,black}&Flat, \SI{350}{\depth}\\
  \entry{solid,blue}&Curved\\
  \entry{solid,red}&Adjusted ($\dagger$)\\
}
\end{tikzpicture}
\caption{\label{fig:cherenkov}
The Cherenkov light yield in arbitrary units (left) and the Cherenkov angle (right) as a function of the inclination angle.
The curves are calculated for the curved atmosphere (blue) and the adjusted atmosphere (red) for an atmospheric slant depth of \SI{350}{\depth} (solid), \SI{435}{\depth}~(dashed) and \SI{520}{\depth} (dotted). They are in good agreement up to at least \SI{70}{\degree}. For reference, also the flat barometric atmosphere is shown (black, dashed) for \SI{350}{\depth}.}
\end{figure}

In Fig.~\ref{fig:cherenkov}, the Cherenkov yield and angle is shown as a function of atmospheric slant depth.

%\afterpage{\clearpage}

\subsection{Shower core extension}\label{sec:shower-radius}

According to the NKG formula~\cite[e.~g.][]{bib:gaisser}, the radius of the shower is given as the Molière radius~$r_\text{M}$ which defines the radius of a cylinder containing on average \SI{90}{\percent} of the shower's energy deposition~\cite{bib:grupen}.
\[\label{eq:molier-radius}
r_\text{M}(h) = \frac{r_\text{s}}{\rho(h)}\]

where $r_\text{s}$ is the Molière unit $X_0\cdot E_\text{s}/E_\text{c}\approx$\,\SI{9.2}{\depth} and $\rho(h=H_\text{max})$ the density of the air at height $h$ of the shower above ground.\\

\begin{figure}
\centering
\begin{tikzpicture}
\footnotesize\sffamily
\image{\includegraphics[width=.495\textwidth]{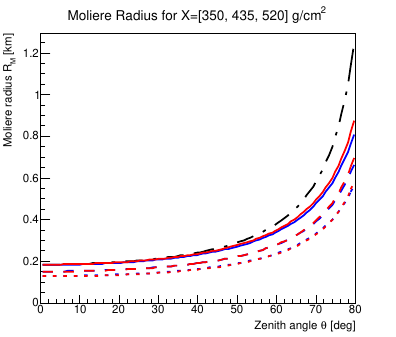}}
\legend{1.05, 4}{
  \entry{dashdotted,black}&Flat, \SI{350}{\depth}\\
  \entry{solid,blue}&Curved\\
  \entry{solid,red}&Adjusted\\
}
\end{tikzpicture}
\begin{tikzpicture}
\footnotesize\sffamily
\image{\includegraphics[width=.495\textwidth]{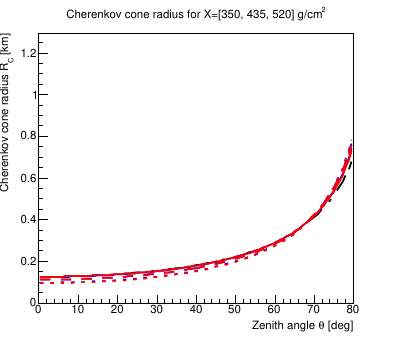}}
\legend{1., 4}{
  \entry{dashdotted,black}&Flat, \SI{350}{\depth}\\
  \entry{solid,blue}&Curved\\
  \entry{solid,red}&Adjusted\\
}
\end{tikzpicture}
\caption{\label{fig:moliere}
The Molière radius (left) as a function of zenith angle and for comparison the radius of the Cherenkov cone on the aperture plane (right). The curves are calculated for the curved atmosphere~(blue) and the adjusted atmosphere (red) for an atmospheric slant depth of \SI{350}{\depth}~(solid), \SI{435}{\depth} (dashed) and \SI{520}{\depth} (dotted). They are in good agreement at least up to \SI{70}{\degree}. For reference, also the flat barometric atmosphere is shown (black, dashed) for \SI{350}{\depth}.}
\end{figure}

The Molière radius is shown in Fig.~\ref{fig:moliere} (left) as a function of zenith angle. For comparison, also the Cherenkov cone at the aperture plane is shown (right).

%\newpage
%\afterpage{\clearpage}

\begin{figure}
\centering
\begin{tikzpicture}
\footnotesize\sffamily
\image{\includegraphics[width=.495\textwidth]{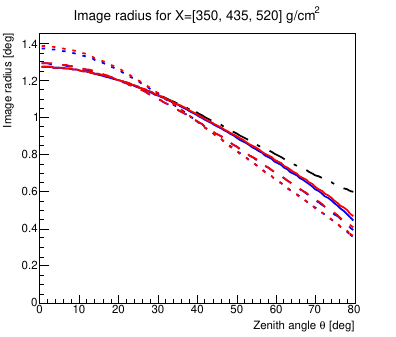}}
\legend{1.1, 1.0}{
  \entry{dashdotted,black}&Flat, \SI{350}{\depth}\\
  \entry{solid,blue}&Curved\\
  \entry{solid,red}&Adjusted ($*$)\\
}
\end{tikzpicture}
\begin{tikzpicture}
\footnotesize\sffamily
\image{\includegraphics[width=.495\textwidth]{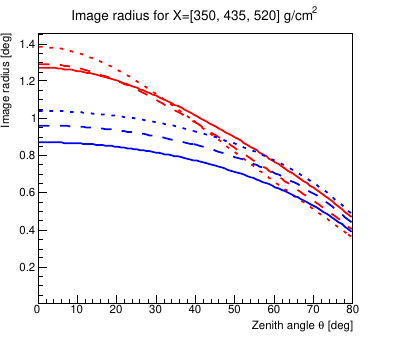}}
\legend{1.1, 1.0}{
  \entry{solid,red}&Image radius ($*$)\\
  \entry{solid,blue}&Cherenkov angle ($\dagger$)\\
}
\end{tikzpicture}
\caption{\label{fig:image-size}The radius of the image size of the shower core as a function of zenith angle (left) and the same (black) compared with the size of the Cherenkov angle (blue) on the right. The left curves are calculated for the curved atmosphere (blue) and the adjusted atmosphere (red) for an atmospheric slant depth of \SI{350}{\depth} (solid), \SI{435}{\depth} (dashed) and \SI{520}{\depth} (dotted). They are in good agreement at least up to \SI{70}{\degree}. For reference, also the flat barometric atmosphere is shown (black, dashed) for \SI{350}{\depth}.}
\end{figure}
%\afterpage{\clearpage}

\subsection{Changes with zenith angle}

In the following, several effects are discussed. For all these effects, the assumption is made that all the light originates from the shower core. Since showers are transparent, this is generally not true. Consequently, each of the described effect should be considered as an approximation. In particular, the longitudinal development of the shower is neglected. This is justified by the small opening angle of Cherenkov light. Showers are mainly viewed along the shower axis or with a small angle to the axis. Thus an elongation of the shower might change the asymmetry in the image but would usually not significantly alter the image shape.

%%%%%%%%%%%%%%%%%%%%%%%%%%%%%%%%%%%%%%%%%%%%%%%% new %%%%%%%%%%%%%%%%%%%%%%%%%%%%%%%%%%%%%%%%%%%%%%

\subsubsection{Cherenkov light yield}\label{sec:light-yield}

The Cherenkov light yield $Y$ is changing with the atmospheric depth of the shower core in the atmosphere and the corresponding refractive index.
\[Y(n)\equiv \frac{\dd N}{\dd x} \propto 1-\frac{1}{n^2}\]

As a change in light yield alters the energy corresponding to the same measured signal, for a power law spectrum, the measured change in trigger rate is well described by the change of the lower energy threshold of the detector. This energy threshold is inversely proportional to the Cherenkov light yield. This is defined in the following as $\nu(E, \theta)$
\[
\frac{E_\text{th}(\theta)}{E_\text{th}(0\degree)}
\propto
\frac{Y(0\degree)}{Y(\theta)}
\equiv
\nu(E, \theta)
\]

The exact solution is then given as
\[\label{eq:nu}
\nu(E, \theta)
=
\frac{1-\frac{1}{n(0\degree)^2}}{1-\frac{1}{n(\theta)^2}}
=
\left[\frac{n(\theta)}{n(0\degree)}\right]^2\cdot
\left[\frac{n(0\degree)+1}{n(\theta)+1}\right]\cdot
\left[\frac{n(0\degree)-1}{n(\theta)-1}\right]
\]

\paragraph{Approximation} As $n\sim1$ with a precision of $\Delta n<1e-3$, the ratio $n(\theta)/n(0\degree)$ and \mbox{$(n(0\degree)+1)/(n(\theta)+1)$} can be approximated as 1 and the relations from Sec.~\ref{sec:refractive-index}, $n$ can be expressed as
\[
\nu(E, \theta)
\approx
\frac{n(0\degree)-1}{n(\theta)-1} 
%= 
%\frac{f(0\degree)}{f(\theta)}
\]

For gases, $n-1$ is proportional to the density of the gas as long as the chemical composition does not change~\cite{bib:stone} which yields
\[
\nu(E, \theta)
\approx
\frac{\rho(0\degree)}{\rho(\theta)}
%=
%\frac{\text{e}^{-c\cdot h(0\degree)^k}}{\text{e}^{-c\cdot h(\theta)^k}}
\]

Assuming a flat atmosphere and an exponential density profile of the atmosphere (barometric atmosphere) with a scale length $H_0$ and applying Eq.~\ref{eq:height-flat} gives%\todo{Compare with non exponential profile! Is there maybe a trivial solution for it?}
\[\label{eq:nu-approx}
\nu(E, \theta)
\approx
\frac
{\text{e}^{-\frac{H_\text{max}(E, 0\degree)}{H_0}}}
{\text{e}^{-\frac{H_\text{max}(E, \theta)}{H_0}}}
=
\frac
{\text{e}^{\frac{h_0}{H_0}\ln\frac{X_\text{max}(E)}{X_0}}}
{\text{e}^{\frac{h_0}{H_0}\ln\frac{X_\text{max}(E)\cdot\cos\theta}{X_0}}}
=
(\cos\theta)^{-\chi}
\]

\noindent with the first of the three fit-parameters 
\[\chi\equiv\frac{h_0}{H_0}\label{eq:chi}\]

Since the scale length for the density profile of the atmosphere directly leads to the scale length of the atmospheric depth, generally the assumption $\chi\equiv1$ is valid. Including effects of the curved atmosphere and a non-exponential profile, the coefficient $\chi$ changed. A good agreement is obtained with $\chi\equiv0.9$.\\

For the adjusted atmosphere, the same would lead to
\[
\nu(E, \theta)
\approx
\frac
{\text{e}^{\frac{h_0}{H_0}\ln\frac{X_\text{max}(E)}{X_0}}}
{\text{e}^{\frac{h_0}{H_0}\left(\cos\theta\right)^\alpha\ln\frac{X_\text{max}(E)\cdot\cos\theta}{X_0}}}
=
\left(\frac{X_\text{max}(E)}{X_0}\right)^{\chi-\chi\cdot\left(\cos\theta\right)^\alpha}\left(\cos\theta\right)^{-\chi\left(\cos\theta\right)^\alpha}
\]

Although, this is more accurate, the energy dependence, makes this form generally less useful. As the difference is only a few percent which can mainly be accounted for by an adapted coefficient~$\chi$, this form will not be used further on.\\

\begin{figure}
\centering
\begin{tikzpicture}
\footnotesize\sffamily
\image{\includegraphics[width=.495\textwidth]{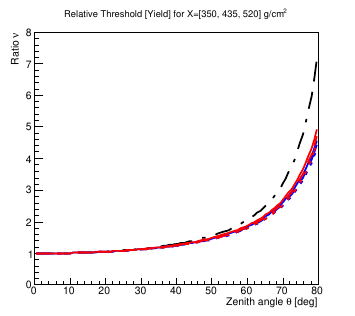}}
\legend{1.1, 4.3}{
  \entry{dashdotted,black}&Flat\\
  \entry{solid,blue}&Curved\\
  \entry{solid,red}&Adjusted ($*$)\\
}
\end{tikzpicture}
\begin{tikzpicture}
\footnotesize\sffamily
\image{\includegraphics[width=.495\textwidth]{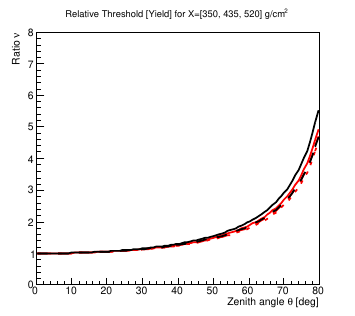}}
\legend{1.1, 4.3}{
  \entry{solid,red}&Adjusted atmosphere($*$)\\
  \entry{solid,black}&Approximation, \(\chi\equiv1\)\\
  \entry{dashed,black}&Approximation, \(\chi\equiv0.9\)\\
}
\end{tikzpicture}
\caption{\label{fig:nu-theta}
The coefficient $\nu$ as a function of zenith angle. The left curves are calculated for the curved (blue) and the adjusted atmosphere (red) for an atmospheric slant depth of \SI{350}{\depth} (solid), \SI{435}{\depth} (dashed) and \SI{520}{\depth} (dotted). All curved are very similar. For reference, also the flat atmosphere is shown (black, solid) for \SI{350}{\depth}. On the right, the numerical calculation (red) is compared with the approximation (black) from Eq.~\ref{eq:nu-approx} with $\chi\equiv1$ (solid) and $\chi\equiv0.9$ (dashed). A good agreement is visible for $\chi\equiv0.9$.}
\end{figure}

In Fig.~\ref{fig:nu-theta}, the coefficient $\nu$ is shown. It can be seen that the ratio is nearly independent of the atmospheric depth and thus primary energy of the particle. 

\subsubsection{Absorption in the atmosphere}\label{sec:absorption}

The mass between the shower core and the telescope is increasing with increasing zenith angle. Consequently, light absorption must increase as well. In a first-order approximation it is assumed that light absorption is proportional to the air-density. Different absorption effects as they take place in the atmosphere are neglected. Certainly, taking them into account a very precise model can be obtained which is not the primary goal of this paper.

The amount of penetrated matter is directly proportional to the optical depth between the shower core and the telescope. Therefore, the change of light intensity $I$ at the aperture plane must be proportional to
\[I(E, \theta) \propto \text{e}^{-\frac{X_\text{tot}(\cos\theta)-X_\text{max}(E)}{x_0}}\]

with the absorption depth $x_0$. 

Since the energy threshold is inversely proportional to that intensity, the corresponding coefficient describing the change in energy is defined as $\eta(E, \theta)$ as
\[
\frac{E_\text{th}(\theta)}{E_\text{th}(0\degree)}
\propto
\frac{I(0\degree)}{I(\theta)}
\equiv
\eta(E, \theta)
\]

The coefficient is then given by
\[\label{eq:eta}
\eta(E, \theta)
=
\frac
{\text{e}^\frac{X_\text{tot}(\cos\theta)-X_\text{max}(E)}{x_0}}
{\text{e}^\frac{X_\text{tot}(0\degree)-X_\text{max}(E)}{x_0}}
=
\text{e}^\frac{X_\text{tot}(\cos\theta)-X_0}{x_0}
=
\left(\text{e}^{1-\frac{1}{\cos\theta}}\right)^{-\xi}
\]

\noindent with the second of the three fit parameters
\[\xi\equiv\frac{X_0}{x_0}\label{eq:xi}\]

This is an exact solution and independent of the energy. Thus
\[\eta(E, \theta) \equiv \eta(\theta)\]

The coefficient $\xi$ is a measure for the absorption in the atmosphere. Strictly speaking, this absorption is dependent on many effects, the coefficient $\xi$ might depend on height and thus indirectly on energy. Generally, speaking the atmosphere is transparent ($x_0>X_0$) and thus $\xi<1$. It is remarkable that it does not depend on $X_\text{max}$ itself but only on the calculation of the total depth of the atmosphere along the line-of-sight. \\

\begin{figure}
\centering
\begin{tikzpicture}
\footnotesize\sffamily
\image{\includegraphics[width=.495\textwidth]{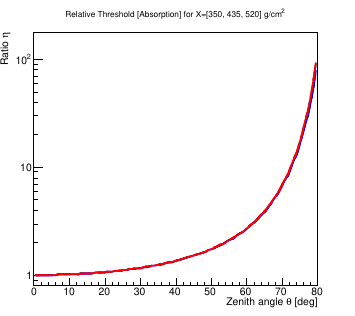}}
\legend{1.1, 4.2}{
  \entry{dashdotted,black}&Flat, \SI{350}{\depth}\\
  \entry{solid,blue}&Curved\\
  \entry{solid,red}&Adjusted ($*$)\\
}
\end{tikzpicture}
\begin{tikzpicture}
\footnotesize\sffamily
\image{\includegraphics[width=.495\textwidth]{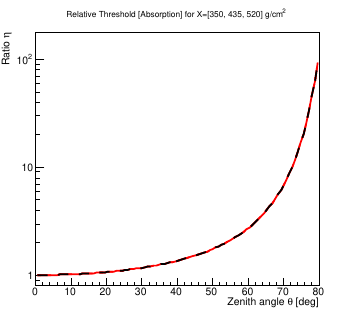}}
\legend{1.1, 4.7}{
  \entry{solid,red}&Adjusted atmosphere ($*$)\\
  \entry{dashed,black}&Simplified, \(\xi\equiv1\)\\
}
\end{tikzpicture}
\caption{\label{fig:eta-theta}
The coefficient $\eta$ as a function of zenith angle. The left curves are calculated for the curved (blue) and the adjusted atmosphere (red) for an atmospheric slant depth of \SI{350}{\depth}~(solid), \SI{435}{\depth} (dashed) and \SI{520}{\depth} (dotted). They almost perfectly overlay each other. For reference, also the flat atmosphere is shown (black, dashed) for \SI{350}{\depth} but hidden below the other curves. On the right, the simplified solution from Eq.~\ref{eq:eta} is shown (black, dashed) for $\xi\equiv1$. This is compared with the adjusted atmosphere (red). The perfect match is expected just plotting two different representations of the same formula. It is, nevertheless, kept for consistence with the other coefficients. All curved are plotted for $\xi\equiv1$.}
\end{figure}

The coefficient $\eta(\theta)$ is shown in Fig.~\ref{fig:eta-theta} for $\xi\equiv1$.

%\afterpage{\clearpage}

%The coefficient $\eta(\theta)$ is shown in Fig.~\ref{fig:eta-theta} for $\xi\equiv1$. 
% \begin{figure}
% \centering
% \includegraphics[width=.495\textwidth]{img/plot-Epsilon-Theta}
% \includegraphics[width=.495\textwidth]{img/plot-Epsilon-Theta-Approx}
% \caption{\label{fig:epsilon-theta}
% The coefficient $\epsilon$ as a function of zenith angle. The left curves are calculated for the Curved (blue) and the Adjusted Atmosphere (red) assuming for an atmospheric slant depth of 350\,g/cm$^2$ (solid), 435\,g/cm$^2$ (dashed) and 520\,g/cm$^2$ (dotted). For reference, also the Flat Atmosphere is shown (black, dashed) for 350\,g/cm$^2$. On the right, the simplified solution is shown for $\chi\equiv1$ (black, solid) and $\chi\equiv0.935$ (black, dashed). This is compared with the Adjusted Atmosphere (red).}
% \end{figure}
%\includegraphics[width=.68\textwidth]{img/plot-Eta-Theta.png}
%\caption{\label{fig:eta-theta}The coefficient $\eta$ as a function of the line-of-sight with an inclination $\theta$. The coefficient is not dependent on the atmospheric model. The adjusted (red) and flat barometric atmosphere (black) is shown for consistency with the other plots.}
%\afterpage{\clearpage}

\subsubsection{Photon density at aperture}\label{sec:aperture-density}

With increasing zenith angle, the shower core moves further away from the telescope and higher above ground. This leads to an increase of the geometrical shower radius (see Eq.~\ref{eq:molier-radius}) and an increase of the emission angle of the light (see Eq.~\ref{eq:cangle}).

Since the size of the light pool on aperture plane is dominated by the deviation of the particles from the shower axis over the emission angle of Cherenkov light, an increase of the emission angle might lead to a change of the steepness of the edges of the light pool on the aperture plane but does, in a first order, not change the size of the illuminated area. The illuminated area instead depends only on the size of the shower core itself. A larger geometrical extension of the shower therefore leads to a decrease of the light density at the aperture and, consequently, an increase of the energy threshold.

The corresponding coefficient $\epsilon(E, \theta)$ is defined as
\[
\frac{E_\text{th}(\theta)}{E_\text{th}(0\degree)}
\propto
\frac{A_\text{M}(E, \theta)}{A_\text{M}(E, 0\degree)}
\equiv
\epsilon(E, \theta)
\]

\noindent and can be derived as
\[\label{eq:epsilon}
\epsilon(E, \theta)
=
\left(
\frac
{r_\text{M}(E, \theta)}
{r_\text{M}(E, 0\degree)}
\right)^2
=
\left(
\frac
{\rho(H_\text{max}(E, 0\degree))}
{\rho(H_\text{max}(E, \theta))}
\right)^2
\]

\paragraph{Approximation} In analogy to Eq.~\ref{eq:nu-approx}, for a flat barometric atmosphere, this can be expressed as%\todo{Compare with non exponential profile! Is there maybe a trivial solution for it?}
\[\label{eq:epsilon-approx}
\epsilon(E, \theta)
\approx
(\cos\theta)^{-2\chi}
\]

\begin{figure}
\centering
\begin{tikzpicture}
\footnotesize\sffamily
\image{\includegraphics[width=.495\textwidth]{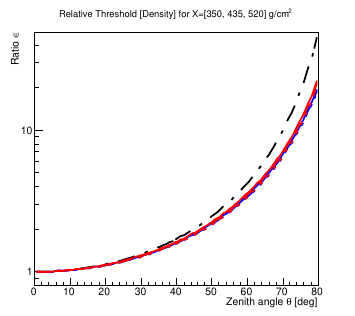}}
\legend{1.1, 4.2}{
  \entry{dashdotted,black}&Flat, \SI{350}{\depth}\\
  \entry{solid,blue}&Curved\\
  \entry{solid,red}&Adjusted ($*$)\\
}
\end{tikzpicture}
\begin{tikzpicture}
\footnotesize\sffamily
\image{\includegraphics[width=.495\textwidth]{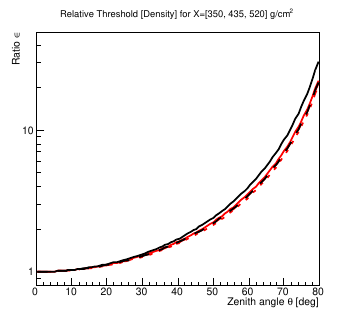}}
\legend{1.1, 4.2}{
  \entry{solid,red}&Adjusted atmosphere ($*$)\\
  \entry{solid,black}&Approximation, \(\chi\equiv1\)\\
  \entry{dashed,black}&Approximation, \(\chi\equiv0.9\)\\
}
\end{tikzpicture}
\caption{\label{fig:epsilon-theta}
The coefficient $\epsilon$ as a function of zenith angle. The left curves are calculated for the curved (blue) and the adjusted atmosphere (red) for an atmospheric slant depth of \SI{350}{\depth}~(solid), \SI{435}{\depth} (dashed) and \SI{520}{\depth} (dotted). They are in good agreement. For reference, also the flat atmosphere is shown (black, dashed) for \SI{350}{\depth}. On the right, the three curves for the adjusted atmosphere (red) are compared with the approximation (black) from Eq.~\ref{eq:epsilon-approx} for $\chi\equiv1$~(solid) and $\chi\equiv0.9$ (dashed). A good agreement is visible for $\chi\equiv0.9$.}

%The coefficient $\epsilon$ as a function of the inclination $\theta$. The solution for the adjusted atmosphere (Eq.~\ref{eq:epsilon}) is shown in red, the flat barometric approximation (Eq.~\ref{eq:epsilon-approx}) in blue. The red curve is calculated for an atmospheric depth of \SI{350}{\depth} (solid), \SI{435}{\depth} (dashed) and \SI{520}{\depth} (dotted). For reference, also the flat barometric atmosphere is shown (black, dashed) for \SI{350}{\depth}. On the right, the simplified solution is shown for $\chi\equiv1$ (black, solid) and $\chi\equiv0.935$ (black, dashed). This is compared with the adjusted atmosphere (red).}
\end{figure}

The coefficient $\epsilon(X, \theta)$ is shown in Fig.~\ref{fig:epsilon-theta}. The numerical solution for Eq.~\ref{eq:epsilon} is compared with the approximation (Eq.~\ref{eq:epsilon-approx}, black) assuming $\chi\equiv1$. A slightly better agreement is obtained with $\chi\equiv0.935$. It is apparent that the result is almost independent of the primary energy and thus the atmospheric depth of the shower maximum.\\

%Assuming a flat barometric atmosphere, i.~e.\ an exponential density profile with a scale length $h_0$ expressing $H_\text{max}$ as a function of the atmospheric depth (Eq.~\ref{eq:height-flat}) yields\todo{Compare with non exponential profile! Is there maybe a trivial solution for it?}

%\[\label{eq:epsilon-approx}
%\epsilon(E, \theta)
%\approx
%\left(
%\frac
%{\text{e}^{-\frac{H_\text{max}(E, 0\degree)}{h_0}}}
%{\text{e}^{-\frac{H_\text{max}(E, \theta)}{h_0}}}
%\right)^2
% %
% =
% \left(
% \frac
% {\text{e}^{\frac{H_0}{h_0}\ln\frac{X_\text{max}(E)}{X_\text{tot}(0\degree)}}}
% {\text{e}^{\frac{H_0}{h_0}\ln\frac{X_\text{max}(E)}{X_\text{tot}(\theta)}}}
% \right)^2
% %
% =
% \left(
% \frac{X_\text{tot}(\theta)}{X_\text{tot}(0\degree)}
% \right)^{2\chi}
% %
% =
% (\cos\theta)^{-2\chi}
% \]

% with 

% \[\chi\equiv\frac{H_0}{h_0}\]

The coefficient $\epsilon(X, \theta)$ is shown in Fig.~\ref{fig:epsilon-theta}. The numerical calculation for the adjusted atmosphere (Eq.~\ref{eq:epsilon}, red) is compared with the flat barometric approximation from Eq.~\ref{eq:epsilon-approx} (black) assuming $\chi\equiv1$. A slightly better agreement is obtained with $\chi=0.9$. It is apparent that the result is almost independent of the primary energy, i.e. the atmospheric depth of the shower maximum. 

% \begin{figure}
% \centering
% \includegraphics[width=.68\textwidth]{img/plot-Epsilon-Theta.png}
% \caption{\label{fig:epsilon-theta}}
% \end{figure}
% \afterpage{\clearpage}

\subsubsection{Photon density at the focal plane}

The photons arriving at the aperture are focused by the optical system on the focal plane where they are detected with photon sensors like semi-conductors (SiPM) or photo-multipliers. A trigger decision is then taken according to the signal in one or more of these sensors which can be understood as a requirement on the minimum photon density in the focal plane.

A shower core which moves further away from the telescope creates a smaller image on the focal plane. Assuming the same photon density at the aperture, that means that they are focused into a smaller area on the camera and thus increase the photon density of the image. This is a pure geometrical effect and can be treated independently of the synchronous decrease of the photon density at the aperture plane due to the increased distance. Consequently, the energy threshold is inversely proportional to the image size. A scale for the image size can be defined by the angle $\delta$ at which the core is seen from the telescope. It should be mentioned that in fact not only the width of the shower core but also the length of the shower determined the image size. Nevertheless, the given description is a good first order approximation. The real effect might be stronger or weaker. Since shower images are small with typical sizes of less than 1\textdegree{}, a small angle approximation can be made.%\todo{plot}
\[\omega(E, \theta) = 
\atan\frac{r_{M}(E, \theta)}{D_\text{max}(E, \theta)}\approx
\frac{r_{M}(E, \theta)}{D_\text{max}(E, \theta)}\]

The relative change of threshold is then defined as the coefficient $\delta(E, \theta)$
\[
\frac{E_\text{th}(\theta)}{E_\text{th}(0\degree)}
\propto
\left(\frac{\omega(E, \theta)}{\omega(E, 0\degree)}\right)^2
\equiv
\delta(E, \theta)
\]

It is then obtained as
\[\label{eq:delta}
\delta(E, \theta)
=
\left(
\frac
{\atan\frac{r_{M}(E, \theta)}{D_\text{max}(E, \theta)}}
{\atan\frac{r_{M}(E, 0\degree)}{D_\text{max}(E, 0\degree)}}
\right)^2
\]

\paragraph{Approximation} This can be approximated as
\[
\delta(E, \theta)
\approx
\left(
\frac
{r_{M}(E, \theta)} 
{r_{M}(E, 0\degree)}
\cdot
\frac
{D_\text{max}(E, 0\degree)}
{D_\text{max}(E, \theta)}
\right)^2
\]

Furthermore, the approximation from Sec.~\ref{sec:aperture-density} can be used. If $D_\text{max}$ is expressed by \mbox{$H_\text{max}=D_\text{max}\cdot\cos\theta$}, another term $\cos^2\theta$ arises which yields
\[
\delta(E, \theta)
\approx
(\cos\theta)^{-2\chi}
\left(
\frac
{D_\text{max}(E, 0\degree)}
{D_\text{max}(E, \theta)}
\right)^2
=
(\cos\theta)^{2(1-\chi)}
\left(
\frac
{H_\text{max}(E, 0\degree)}
{H_\text{max}(E, \theta)}
\right)^2
\]
%%%%%%%%%%%%%%%%%%%%%%%%%%%%%%%%%%%%%% new %%%%%%%%%%%%%%%%%%%%%%%%%%%%%%%%%%%%%%%%
The ratio in $H_\text{max}$ can now be expressed applying the adjusted atmosphere.
\[
\frac
{H_\text{max}(E, 0\degree)}
{H_\text{max}(E, \theta)}
=
\frac
{\ln\frac{X_\text{max}(E)}{X_\text{0}(0\degree)}}
{\left(\cos\theta\right)^\alpha\cdot\ln\frac{X_\text{max}(E)}{X_\text{0}(\theta)}}
=
\left(\cos\theta\right)^{-\alpha}
\frac
{\ln\frac{X_\text{max}(E)}{X_0}}
{\ln\frac{X_\text{max}(E)}{X_0}+\ln\cos\theta}
=
\left(\cos\theta\right)^{-\alpha}
\frac
{1}
{1-\kappa\ln\cos\theta}
\]

\noindent with the third of the three fit parameters $\kappa$ defined as
\[\kappa(E)\equiv\left(-\ln\frac{X_\text{max}(E)}{X_0}\right)^{-1}\label{eq:kappa}\]

For typical $X_\text{max}(E)$ between 350\,g/cm$^2$ and 520\,g/cm$^2$, $\kappa(E)$ varies between 0.93 and 1.45. It replaces the energy dependence in the approximation.

Because $(\ln\cos\theta)^2\ll\ln\cos\theta$ and $|\ln\cos\theta|<1$, this can further be approximated using a binomial series
\[(1-\ln\cos\theta)^\kappa = \sum\limits_{i=0}^\infty \binom{\kappa}{i} (-\ln\cos\theta)^i \approx 1 - \kappa\cdot \ln\cos\theta\]

\noindent to 
\[\label{eq:delta-approx}
\delta(E, \theta)
\equiv
\delta(\theta)
\approx
(\cos\theta)^{2(1-\chi-\alpha)}
\left(1 - \ln\cos\theta\right)^{-2\kappa}
\]

Due to this approximation, it is expected that $\kappa(E)$ is slightly altered compared with the ideal value.\\

\begin{figure}
\centering
\begin{tikzpicture}
\footnotesize\sffamily
\image{\includegraphics[width=.495\textwidth]{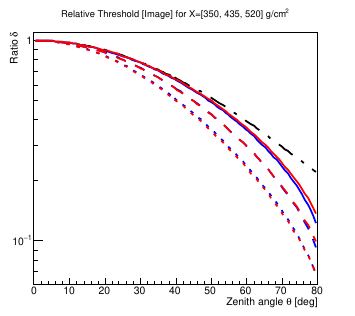}}
\legend{1.1, 1.1}{
  \entry{dashdotted,black}&Flat, \SI{350}{\depth}\\
  \entry{solid,blue}&Curved\\
  \entry{solid,red}&Adjusted ($*$)\\
}
\end{tikzpicture}
\begin{tikzpicture}
\footnotesize\sffamily
\image{\includegraphics[width=.495\textwidth]{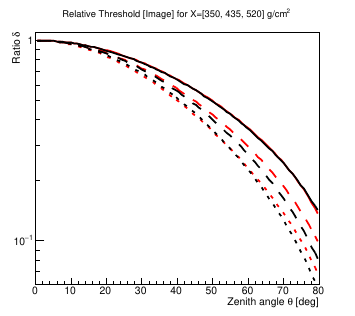}}
\legend{1.05, 1.0}{
  \entry{solid,red}&Adjusted atmosphere ($*$)\\
  \entry{solid,black}&Approximation, \(\kappa\equiv0.93\)\\
  \entry{dashed,black}&Approximation, \(\kappa\equiv1.15\)\\
  \entry{dotted,black}&Approximation, \(\kappa\equiv1.45\)\\
}
\end{tikzpicture}
\caption{\label{fig:delta-theta}
The coefficient $\delta$ as a function of zenith angle. The left curves are calculated for the curved (blue) and the adjusted atmosphere (red) assuming for an atmospheric slant depth of \SI{350}{\depth} (solid), \SI{435}{\depth} (dashed) and \SI{520}{\depth} (dotted). They show a good agreement. For reference, also the flat atmosphere is shown (black, dashed) for \SI{350}{\depth}. On the right, the three results for the adjusted atmosphere (red) are compared with the approximation (black) from Eq.~\ref{eq:delta-approx} for $\chi\equiv0.9$. A reasonable match is obtained for $\kappa\equiv0.93$ (solid), $\kappa\equiv1.15$ (dashed) and $\kappa=1.45$ (dotted) which corresponds to an atmospheric depth of \SI{350}{\depth}, \SI{435}{\depth} and \SI{520}{\depth}, respectively. A reasonable agreement is visible.}
\end{figure}

The coefficient $\delta(E, \theta)$ is shown in Fig.~\ref{fig:delta-theta}. The calculation for the adjusted atmosphere~(Eq.~\ref{eq:delta}, red) is compared with the flat barometric approximation (Eq.~\ref{eq:delta-approx}, black).

\subsubsection{Field of view}

%A shower core is only visible in the field-of-view of the camera if its light illuminates the optics of the telescope. This effect is included in the collection area. 

The detected solid angle is the field-of-view of the the camera of the telescope system and thus constant.
\[\Omega(E, \theta)\approx 1\]

\subsubsection{Other effects}

Other effect or a more precise description of the discussed effect might alter the result. However, it is fair to assume that a more precise description will mainly alter the strength of the effects. This can usually be expressed with a higher order polynomial and thus be approximated by a change of the exponent. Therefore, it is not expected that the analytical shape of the result changes b or this change can be expressed by an artificially introduced exponent.

%\newpage
\subsection{Summary}

The change in rate with zenith angle according to Eq.~\ref{eq:rate} is given by
\[\label{eq:ratio}
\frac{R(\theta)}{R(0\degree)} = 
\frac
{\int\limits_{E_\text{th}(\theta)}^\infty \phi(E)\cdot A_\text{eff}(E, \theta) \,\dd{E}}
{\int\limits_{E_\text{th}(0 \degree)}^\infty \phi(E)\cdot A_\text{eff}(E, 0\degree) \,\dd{E}} \]

In this formula for the rate-change, the energy threshold $E_\text{th}$ and the effective collection area~$A_\text{eff}$ are dependent on the zenith angle $\theta$ of observation. The primary spectrum $\Phi$ is independent of the viewing angle.

\paragraph{Energy Threshold} As discussed on the previous sections, the energy threshold is proportional to

\begin{itemize}
\item $\nu(\theta)$ due to changes in the Cherenkov light yield as a function of the atmosphere's density profile,
\item $\eta(\theta)$ due to the change in photon density at the aperture caused by the change in atmospheric absorption,
\item $\epsilon(\theta)$ due to the change in photon density at the aperture caused by the geometrical effects, and
\item $\delta(\theta)$ due to the change in photon density at the focal plane caused by the change in image size.
\end{itemize}

Therefore, the threshold at a given zenith angle $\theta$ can be expressed as
\[\label{eq:coefficients}
E_\text{th}(\theta) \approx [\nu(\theta)\cdot\epsilon(\theta)\cdot\eta(\theta)\cdot\delta(\theta)]\cdot E_\text{th}(\degree)\]

\begin{figure}
\centering
\begin{tikzpicture}
\footnotesize\sffamily
\image{\includegraphics[width=.495\textwidth]{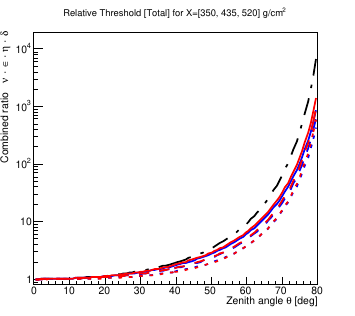}}
\legend{1.05, 4.3}{
  \entry{dashdotted,black}&Flat, \SI{350}{\depth}\\
  \entry{blue,solid}&Curved\\
  \entry{red,solid}&Adjusted ($*$)\\
}
\end{tikzpicture}
\begin{tikzpicture}
\footnotesize\sffamily
\image{\includegraphics[width=.495\textwidth]{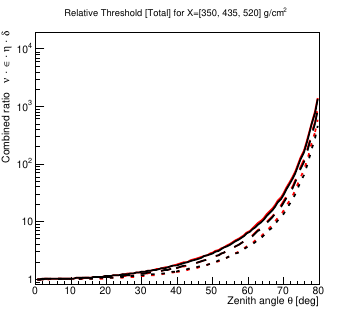}}
\legend{1.05, 3.9}{
  \entry{solid,red}&Adjusted atmosphere ($*$)\\
  \entry{solid,black}&Approximation, \(\kappa\equiv0.93\)\\
  \entry{dashed,black}&Approximation, \(\kappa\equiv1.15\)\\
  \entry{dotted,black}&Approximation, \(\kappa\equiv1.45\)\\
}
\end{tikzpicture}
\caption{\label{fig:relth-theta}The product of the coefficients $\chi\cdot\epsilon\cdot\eta\cdot\delta$ as a function of zenith angle. The left curves are calculated for the curved (blue) and the adjusted atmosphere (red) assuming for an atmospheric slant depth of \SI{350}{\depth} (solid), \SI{435}{\depth} (dashed) and \SI{520}{\depth} (dotted). A good agreement is visible. For reference, also the flat atmosphere is shown (black, dashed) for \SI{350}{\depth}. On the right, the approximations (black) from Eqs.~\ref{eq:nu-approx}, \ref{eq:epsilon-approx}, \ref{eq:eta}, \ref{eq:delta-approx} with the coefficients $\chi\equiv0.9$, $\xi\equiv1$, and $\kappa\equiv0.93$ (solid), $\kappa\equiv1.15$ (dashed) and $\kappa=1.45$ (dotted)  corresponding to an atmospheric depth of \SI{350}{\depth}, \SI{435}{\depth} and \SI{520}{\depth}, respectively. This is compared with the three curves for the adjusted atmosphere (red). A good agreement is visible.}
\end{figure}

The product of the coefficients $\nu\cdot\epsilon\cdot\eta\cdot\delta$ is shown in Fig.~\ref{fig:relth-theta}. The numerical solution for Eqs.~\ref{eq:nu}, \ref{eq:epsilon}, \ref{eq:eta}, \ref{eq:delta} applying the adjusted atmosphere (red) is compared with the approximation (Eqs.~\ref{eq:nu-approx}, \ref{eq:epsilon-approx}, \ref{eq:eta}, \ref{eq:delta-approx}). In the following, they are referred to as {\em Solution~A} and {\em Solution~B}, respectively.

% \begin{figure}
% \centering
% \includegraphics[width=.68\textwidth]{img/plot-RelTH-Theta.png}
% \caption{\label{fig:relth-theta}The product of the coefficients $\chi\cdot\epsilon\cdot\eta\cdot\delta$ as a function of the line-of-sight with an inclination $\theta$. The numerical solution for the adjusted atmosphere is shown in red for an atmospheric depth of 350\,g/cm$^2$ (solid), 435\,g/cm$^2$ (dashed) and 520\,g/cm$^2$ (dotted). The approximation is shown in black for $\chi\equiv1$, $\xi\equiv1$ and $\zeta\equiv2.3$.}
% \end{figure}

%\afterpage{\clearpage}

\paragraph{Collection Area} As the area illuminated by a shower suffers an exponential cut-off at the edges, in a first approximation, the effective collection area must be proportional to the illuminated area on the aperture plane and thus 
\[A_\text{eff}(E, \theta) \approx \epsilon(\theta)\cdot A_\text{eff}(E, 0\degree) \]

This allows to rewrite the rate change as
\[\label{eq:full-integral}
\frac{R(\theta)}{R(0\degree)} = 
\epsilon(\theta)\cdot 
\frac
{\int\limits_{E_\text{th}(\theta)}^\infty \phi(E)\cdot A_\text{eff}(E, 0\degree) \,\dd{E}}
{\int\limits_{E_\text{th}(0 \degree)}^\infty \phi(E)\cdot A_\text{eff}(E, 0\degree) \,\dd{E}} \]

It should be noted that this approximation becomes invalid at large effective areas because the angle under which the shower can be seen is limited by the field-of-view of the camera. Therefore, showers seen at an angle larger than the field-of-view will not be imaged into the camera and get lost.

\begin{figure}
\centering
\begin{tikzpicture}
\footnotesize\sffamily
\image{\includegraphics[width=.495\textwidth]{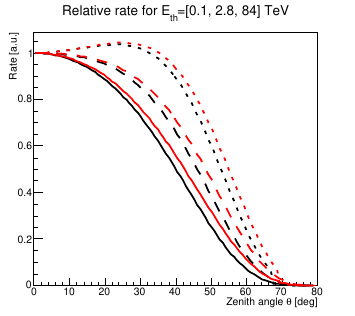}}
\legend{1.05, 1.}{
  \header{Solution A}\\
  \entry{solid,black}&Flat\\
  \entry{solid,red}&Adjusted ($*$)\\
}
\end{tikzpicture}
\begin{tikzpicture}
\footnotesize\sffamily
\image{\includegraphics[width=.495\textwidth]{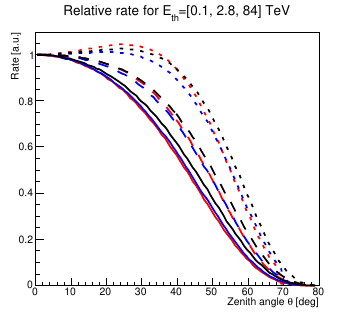}}
\legend{1.05, 1.}{
  \entry{solid,red}&Solution A ($*$)\\
  \entry{solid,blue}&Solution B\\
  \entry{black,solid}&Solution C ($\dagger$)\\
}
\end{tikzpicture}
\caption{\label{fig:rate-theta}The rate in arbitrary units as a function of zenith angle for a spectrum with $\gamma\equiv2.70$. The left curves are calculated numerically for the flat (black) and the adjusted atmosphere (red) from Eq.~\ref{eq:ratio} scaling the cut-off energy $E_\text{th}(0\degree)$ for protons of \SI{100}{GeV} (solid), \SI{2.8}{TeV} (dashed) and \SI{84}{TeV} (dotted) with the coefficients from Eqs.~\ref{eq:nu}, \ref{eq:epsilon}, \ref{eq:eta}, \ref{eq:delta} and $A_\text{eff}\propto r_\text{M}^2$ (Solution A). It shows how the atmosphere influences the zenith-angle cut-off already around \SI{30}{\degree}. 
The numerical solution for the adjusted atmosphere (red) is compared with a numerical calculation using the exact solution for the effective area but the approximations from Eqs.~\ref{eq:nu-approx}, \ref{eq:epsilon-approx}, \ref{eq:eta}, \ref{eq:delta-approx} (Solution B) with the coefficients for $\xi\equiv1$, $\chi\equiv0.9$ and $\kappa\equiv0.93$ (solid), $\kappa\equiv1.15$ (dashed) and $\kappa=1.45$ (dotted) corresponding to an atmospheric depth of \SI{350}{\depth}, \SI{435}{\depth} and \SI{520}{\depth}, respectively. The black lines show the results for the analytical solution (Solution C, Eq.~\ref{eq:final-result}) analogously. All three cases show a reasonably good agreement. Perfect agreement is not expected due to the difference in the atmospheric model and the analytical approximations.}
\end{figure}

\paragraph{Result} The integrals can be solved numerically for the adjusted atmosphere. This is compared with the analytical approximations for the three coefficients to calculate the change in threshold with zenith angle. The result is shown in Fig.~\ref{fig:rate-theta}.\\
%\clearpage

% \begin{figure}
% \centering
% \includegraphics[width=.68\textwidth]{img/plot-Rate-Theta.png}
% \caption{\label{fig:rate-theta}The ratio in trigger rate w.~r.~t.\ the trigger rate at $\theta$=0\textdegree{}. The numerically solution for the adjusted atmosphere is shown in red. The solid, dashed and dotted lines correspond to tree different energy threshold defined by the corresponding atmospheric depth of the shower maximum at \SI{350}{\depth} (solid), \SI{435}{\depth} (dashed) and \SI{520}{\depth} (dotted). The black line shows the same using the simplifications of the coefficients applying a simple barometric, i.e. exponential, profile and a flat atmosphere. The blue line shows the result from the discussed further approximations s summarized in Eq.~\ref{eq:approximation} which are used to obtain a comprehensive analytical expression.}
% \end{figure}
%\afterpage{\clearpage}

Due to the steeply falling spectrum and a slow change of the effective collection area with energy (see also Fig.~\ref{fig:moliere}), in a first order approximation, the effective collection area can be considered energy independent above the energy threshold and thus cancel out from the ratio. In reality, there is a slow nearly linear increase with energy which is neglected here and starts only well above the energy threshold.
\[\label{eq:at-threshold}
\frac{R(\theta)}{R(0\degree)} = 
\epsilon(\theta)\cdot 
\frac
{\int\limits_{E_\text{th}(\theta)}^\infty \phi(E)\,\dd{E}}
{\int\limits_{E_\text{th}(0\degree)}^\infty \phi(E)\,\dd{E}}
=
\epsilon(\theta)\cdot 
\left(\frac{E_\text{th}(\theta)}{E_\text{th}(0\degree)}\right)^{1-\gamma}
=
\epsilon(\theta)\cdot[\nu(\theta)\cdot\epsilon(\theta)\cdot\eta(\theta)\cdot\delta(\theta)]^{1-\gamma}
\]

Replacing $\nu(\theta)$, $\delta(\theta)$, $\eta(\theta)$ and $\epsilon(\theta)$ with the approximations from Eqs.~\ref{eq:nu-approx}, \ref{eq:epsilon-approx}, \ref{eq:eta} and~\ref{eq:delta-approx} yields
\[\label{eq:final-result}
\frac{R(\theta)}{R(0\degree)} = 
\left[\cos\theta^{-2\chi} \right]^{2-\gamma}
\cdot
\left[ 
(\cos\theta)^{-\chi}
\,
\left(\text{e}^{1-\frac{1}{\cos\theta}}\right)^{-\xi}
\,
(\cos\theta)^{2(1-\chi-\alpha)}
(1-\ln\cos\theta)^{-2\kappa}
%\text{e}^{\zeta(\cos\theta-1)} 
\right]^{1-\gamma}
\]

\[\label{eq:final-approx}
= 
\left[\cos\theta\right]^\ell
\cdot
\left[ 
\text{e}^{1-\frac{1}{\cos\theta}} \right]^{\xi(\gamma-1)}
\cdot 
\left[1-\ln\cos\theta\right]^{2\kappa(\gamma-1)}
\]

\noindent with the coefficient $\ell$ defined as
\[
\ell\equiv (\gamma-1)(5\chi+\alpha-2) - 2\chi \approx 1.8\,...\,2.6
\]

The approximate values are given for a spectral index of 2.4\,...\,2.7. \\

Eq.~\ref{eq:final-approx} is referred to as {\em Solution C}.\\
%For the ideal case $\chi\equiv1$ this yields $\ell\equiv3\gamma-5$ and thus for a proton spectrum ($\gamma\sim2.7$) gives $\ell\sim3$

\begin{figure}
\centering
\includegraphics[width=.495\textwidth]{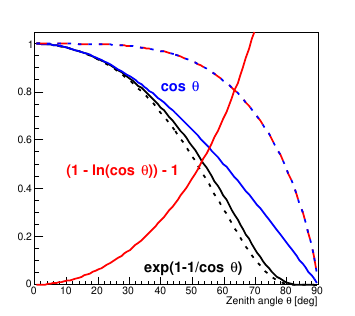}
\caption{\label{fig:terms}The three terms from Eq.~\ref{eq:final-approx} in black ($\exp(1-1/\cos\theta)$), blue ($\cos\theta$) and red ($1-\ln\cos\theta$) shifted by -1. The dotted line is the product of all three. The dashed line (blue-red) is the product $\cos\theta\cdot(1-\ln\cos\theta)$.}
\end{figure}

Fig.~\ref{fig:terms} shows the three terms of Eq.~\ref{eq:final-approx}. While the cut-off and its shape is dominated by the absorption (black), the logarithmic term balances the cosine-term up to medium angles and the tail of the product is thus defined by the ratio of $\chi$ and $\kappa$. This also explains why, if only data up to about 60\textdegree{} is analyzed, only two terms for a reasonable fit are required.

\section{Fitting the data}

The First G-APD Cherenkov Telescope~\cite{bib:design} is the first imaging air-Cherenkov telescope which applied semi-conductor based photo sensors, so-called SiPM, for photo detection. It is dedicated to the long-term monitoring of the brightest blazars~\cite{bib:dwarf,bib:longterm}. Due to the robustness and the high precision of the sensors~\cite{bib:calibration}, the FACT telescope has collected an unprecedented amount of consistent data over many years of data taking. This makes the data ideally suited to investigate efficiency corrections phenomenologically.\\ 

The FACT collaboration has presented a preliminary study on the zenith angle dependence of the trigger rate of an imaging air-Cherenkov telescope~\cite{bib:mahlke} which was the basis for this study. For the presented study, the cosmic ray rates obtained with an artificial software trigger and the gamma-ray excess rates of data taken from the Crab Nebula were presented. For the cosmic ray rates, the threshold of the software trigger was set so high that no dependence on the hardware trigger-threshold was observed anymore. For the Crab Nebula, only good quality data and data with a low trigger-threshold had been used. For the cosmic ray rate, the maximum of the kernel density estimation (KDE) of each zenith angle bin were fitted and for the gamma-ray the sum of the measured events in a given bin divided by the accumulated effective observation time in that bin. For the coefficient $\gamma$, -2.7 was used for the cosmic ray spectrum~\cite{bib:pdl} and -2.49 for the gamma-ray excess rate~\cite{bib:hillas}.\\

\begin{figure}
\centering
\includegraphics[width=.495\textwidth]{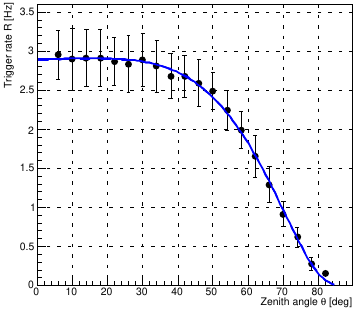}
\hfill
\includegraphics[width=.495\textwidth]{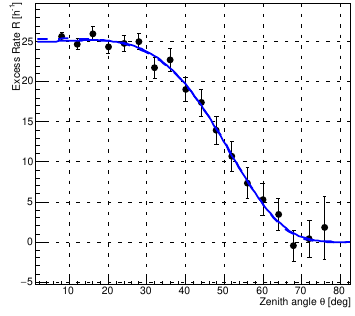}
\caption{\label{fig:final}{\em Left:} Trigger rate as obtained with an artificial software trigger versus zenith angle. The data points are the maxima of the kernel density estimator. The error bars depict the standard deviation of the distribution of the original measurements. {\em Right:} Excess rate as measured for the Crab Nebula versus zenith angle. The error bars are the propagated error from the counting error of the foreground and background measurement. Both measurements are done with the First G-APD Cherenkov Telescope (FACT) and were taken from~\cite{bib:mahlke}. Shown is the fit of the calculation using the exact solution for the effective area but the approximations from Eqs.~\ref{eq:nu-approx}, \ref{eq:epsilon-approx}, \ref{eq:eta}, \ref{eq:delta-approx} (Solution B, dashed) and the final approximation from Eq.~\ref{eq:final-approx} (Solution C, solid). For the results, see Table~\ref{tab:results1} and Table~\ref{tab:results2}, respectively. For the fit of the artificial trigger rate, errors were not taken into account. This explains the small errors on the fit result. The reason is that the shown error bars show the width of the distribution rather than the error on the measurement which must be much smaller due to high statistics. The full numerical solution is not shown as it is graphically indistinguishable from the dashed curve.}
\end{figure}

A fit to this data is shown in Fig.~\ref{fig:final}. It applied the calculation using the exact solution for the effective area but the approximations from Eqs.~\ref{eq:nu-approx}, \ref{eq:epsilon-approx}, \ref{eq:eta}, \ref{eq:delta-approx} (dashed) and the final approximation from Eq.~\ref{eq:final-approx} (solid). For the fit of the background rate, the error bars are ignored as the error on the measurement is much smaller. 

%The energy threshold of the FACT telescope for gammas is in the order of 520\,GeV to 1\,TeV and for the CR rate is in the order of 10\,TeV. This corresponds to a shower maximum between 350\,g/cm$^2$ and 400\,g/cm$^2$ for gammas and around 435\,g/cm$^2$ for protons. 

%\afterpage{\clearpage}

\paragraph{Validity} If the measured rates $R$ are extrapolated to $\theta$=0\textdegree{}, they yield rates of \SI{25}{\per\hour} for Crab and \SI{2.9}{\hertz} for protons, respectively. Approximating the energy threshold of the detector as a sharp cut-off in energy, an effective cut-off energy can be derived from these rates if the measured spectrum is known. The calculated effective spectral cut-off energy $E_{cut}(0\degree)$ for the Crab Nebula is \SI{1.6}{TeV} and \SI{3.3}{TeV} for cosmic rays if a differential energy spectrum of $3.2\cdot10^{-7}\,(E/\text{TeV})^{-2.49}(\text{m}^2\,\text{s}\,\text{TeV})^{-1}$~\cite{bib:hillas} and $1.8\cdot10^{4}\,(E/\text{GeV})^{-2.7}(\text{m}^2\,\text{s}\,\text{sr}\,\text{GeV})^{-1}$~\cite{bib:pdl}, resp., is assumed. These energies are equivalent to an atmospheric depth in the order of \SI{430}{\depth} which converts to a height above ground of the shower core of more than \SI{7}{\kilo\meter} and thus well above ground to see an effect in the data from partial showers reaching out to the detector plane. For large zenith angles around \SI{75}{\degree} it translates to less than \SI{18}{\kilo\meter} above ground (see Fig~\ref{fig:H-X}). An increase of energy threshold of a factor ten is reached around \SI{70}{\degree} according to Fig.~\ref{fig:relth-theta}. For a threshold around \SI{30}{TeV}, $X_\text{max}$ extends to \SIrange{480}{530}{\depth} corresponding to \SIrange{6}{16}{\kilo\meter}. Even for an energy of \SI{100}{TeV} corresponding to an atmospheric depth of \SI{600}{\depth}, the shower core would still be above \SI{4}{\kilo\meter} and the previously discussed assumptions still hold.\\

The mentioned energy cut-off is in good agreement with simulations corresponding to the applied analysis. The shape of the zenith angle dependency fits well with the expectation from Fig.~\ref{fig:rate-theta} for the corresponding atmospheric depth $X_\text{max}\sim$\,\SI{435}{\depth}. It should be noted that these two numbers are not necessarily equivalent to the usual definition of the energy threshold of an imaging air-Cherenkov telescope and the analysis applied to the Crab Nebula in this context was not optimized.

\paragraph{Discussion} The naive expectation for the coefficients $\chi$, $\xi$ and $\kappa$ from the model would be $\chi\equiv1$, $\xi\leq1$ and $\kappa$ between 0.9 and 1.5. Taking the curved atmosphere into account, $\chi\sim0.9$ yielded a better description. 

That the fit does not yield these parameters exactly is expected as they have to correct for the approximations and the assumption of a flat barometric atmosphere in the approximated coefficients. In particular, the description of the atmospheric absorption is over-simplified which needs to be balanced by adapted coefficients. It is also clear that the difference in the elongation rate for gammas and protons, which is totally taken out by the approximations, has a clear influence on the shape of the curve as seen in Fig.~\ref{fig:rate-theta}. Furthermore, the reduction of the shower to a light emitting disc at an atmospheric depth $X_\text{max}$ is a very rough approximation. 

While for the cosmic ray spectrum, the resulting coefficients $\chi$ and $\kappa$ are remarkably close to the expectations, they deviate quite significantly for the gamma-ray rate. Assuming that $\chi=0.9$ corresponding to a reasonable atmospheric model as discussed earlier, is a well-known value and keeping it fixed in the fit, the values for $\kappa$ are roughly consistent with the expected energy range. 

The large deviations for gammas can be understood from the trigger condition. While for the cosmic ray spectrum, the trigger criteria is mainly a simple threshold, for gammas, it includes more complicated steps like image-cleaning and background suppression cuts. While, generally speaking, each of these cuts should show the behavior on the energy threshold as discussed previously, the combination of them can in general alter the zenith angle dependence as they can provide a response differently from image size and brightness such as image shape which is not considered here. Additionally, the applied image-cleaning takes signal timing constraints into account which have not been included in the discussed model at all. Also the assumption that the shower core size dominates over the Cherenkov angle might have to be reconsidered. Another important aspect is that the observed gammas are not diffuse and therefore the effect of losing gammas at high impact parameters due to the limited emission angle of the Cherenkov photons is more pronounced.

%Nevertheless, it is remarkable how well the fit matches the naive expectation in the case of fitting the cosmic ray rate, even better if $\chi$ is fixed to 1.
%The most likely explanation for the larger deviation in case of the gamma-ray rate is that gammas from the source are not diffuse and thus the effect of a cut-off due to the angle at which the shower is observed as discussed in the paragraph ``Collection Area'' is different.

%Instead of the assumption that all the flux is measured at the energy threshold as used for Eq.~\ref{eq:at-threshold}, 
%of an arbitrary proportionality factor, If the energy threshold $E_\text{th}(0\degree)$ at $\theta$=\textdegree{} is kept as a free parameter in the fit, Eq.~\ref{eq:full-integral} can be solved numerically. This leads to a value of $E_\text{th}(0\degree)$ for gammas of 1.6\,TeV and for protons of 3.3\,TeV and a rate of 25/h and 2.9\,Hzm respectively.

%As a fit in which all parameters are free parameters results in 

%\begin{table}
%\begin{tabular}{l|c|c|c}
%    & $\chi$ & $\xi$  & $\zeta$ \\\hline\hline
%Crab   & [0.00] &  1.82  & 2.76\\
%Crab   & [1.00] &  1.59  & 3.12\\
%Crab   &  7.56  &  0.03  & 6.12\\
%Crab   &        & [0.00] &      \\\hline
%CR     &  0.21  &  0.16  & 1.40\\
%CR     & [0.00] &  0.21  & 1.26\\
%CR     &  0.93  & [0.00] & 1.91\\
%CR     & [1.00] &        &     \\
%\end{tabular}
%\caption{\label{tab:results}Result for the coefficients from fitting Eq.~\ref{eq:approximation} to the data. }
%\end{table}

\begin{table}
\centering
\begin{tabular}{|l|c||c|c|c|}\hline
    &&\multicolumn{3}{c|}{Fit results: {\em Solution B}}\\\cline{3-5}
    & $E_\text{TH}$ & $\chi$ & $\xi$ & $\kappa$\\% & $\ell$ \\
  \hline\hline
CR &  \SI{3.33}{TeV} & $0.82\pm0.09$ & $0.21\pm0.11$ & $0.78\pm0.21$\\% & 1.86 \\
CR &  \SI{3.33}{TeV} & 0.9 (fixed)   & $0.12\pm0.02$ & $0.95\pm0.02$\\% & 1.86 \\
\hline
EXC & \SI{1.63}{TeV} & $1.2\pm1.1$   & $1.2\pm1.8$ & $2.2\pm1.9$   \\%& 10.2 \\
EXC & \SI{1.63}{TeV} & 0.9 (fixed)   & $1.8\pm0.4$ & $1.7\pm0.3$   \\% & 1.86 \\
\hline
\end{tabular}
\caption{\label{tab:results1}Results for the coefficients from fitting the exact solution for the effective area but the approximations from Eqs.~\ref{eq:nu-approx}, \ref{eq:epsilon-approx}, \ref{eq:eta}, \ref{eq:delta-approx} (Solution B). The result with a fixed $\chi$ is graphically indistinguishable. CR stands for the fit of the artificial trigger rate which is a measure of the cosmic ray rate. EXC depicts the fit of  the gamma-ray excess rate from the Crab Nebula. The given errors are the errors obtained from the fit.}
\end{table}

\begin{table}
\centering
\begin{tabular}{|l||c|c|c|}\hline
  &\multicolumn{3}{c|}{Fit results: {\em Solution C}}\\\cline{2-4}
  & $\chi$ & $\xi$ & $\kappa$\\% & $\ell$ \\
  \hline\hline
CR &  $0.78\pm0.09$ & $0.26\pm0.08$ & $0.71\pm0.14$\\% & 1.86 \\
CR &  0.9 (fixed)   & $0.15\pm0.02$ & $0.91\pm0.02$\\% & 2.67 \\
\hline
EXC &  $2.4\pm1.4$  & $0.23\pm1.81$ & $3.7\pm1.8$ \\%& 10.2 \\
EXC &  0.9 (fixed)  & $2.2\pm0.4$   & $1.9\pm0.3$ \\%&  2.1 \\
\hline
\end{tabular}
\caption{\label{tab:results2}Result for the coefficients from fitting Eq.~\ref{eq:final-approx} (Solution C) to the data.  The result with a fixed $\chi$ is graphically indistinguishable. CR stands for the fit of the artificial trigger rate which is a measure of the cosmic ray rate. EXC depicts the fit of the gamma-ray excess rate from the Crab Nebula. The given errors are the errors obtained from the fit.}
\end{table}

\section{Conclusions}

From first principles and some approximations, it was possible to obtain an analytical formula which fits the zenith angle dependence of the measured trigger rate for cosmic rays. The formula consists of only three fit-parameters which are linked to the properties of the atmosphere. 

\begin{figure}
\centering
\begin{tikzpicture}
\footnotesize\sffamily
\image{\includegraphics[width=.685\textwidth]{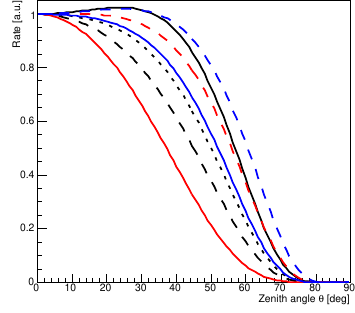}}
\legend{7.05, 8.5}{
  \header{Solution C ($\dagger$)}\\
   \entry{dotted,black}&$\chi\equiv$\,0.9, $\xi\equiv$\,1, $\kappa\equiv$\,1.15\\
}
\legend{7.5, 5}{
   \entry{dashed,blue}&$\xi$ - 0.2\\
   \entry{solid,black}&$\kappa$ + 0.28\\
   \entry{dashed,red}&$\chi$ - 0.2\\
}
\legend{3, 3}{
  \entry{solid,blue}&$\xi$ + 0.2\\
  \entry{dashed,black}&$\kappa$ - 0.22\\
  \entry{solid,red}&$\chi$ + 0.2\\
}
\end{tikzpicture}
\caption{\label{fig:comparison}The rate in arbitrary units as a function of zenith angle for a spectrum with $\gamma\equiv2.70$ for Solution~C (Fig.~\ref{fig:rate-theta}, black dashed). With respect to this curve each of the other curves represent the result changing a single parameter. A solid line always refers to an increase of the parameter, a dashed line to a decrease.}
\end{figure}

The first fit parameter $\chi$~(Eq.~\ref{eq:chi}) is related to deviations of the atmosphere from the simple barometric pressure profile; the second fit parameter $\xi$ (Eq.~\ref{eq:xi}) is proportional to the absorption length of the emitted photons relative to the total overburden of the atmosphere; and the third parameter $\kappa$ (Eq.~\ref{eq:kappa}) is related to the change of the shower core height $X_\text{max}$ with energy, the so-called elongation rate. How a change of each of the three parameter alters the result is shown in Fig.~\ref{fig:comparison}.
For gamma-rays, the provided formula still provides a good description of the data, although the meaning of the obtained parameters has to be taken with care. Even though, the goal of this study was not to obtain meaningful parameters from the fit, it seems likely that with some additional work they can be derived from the fit parameters. This might be useful to obtain a measurement on the atmospheric absorption similar to~\cite{bib:hahn}. In particular for atmospheric absorption, this study made a very rough estimate only which can certainly be improved. Also, the longitudinal development of a shower could be included.

This fit function describes the zenith angle dependence of both, gamma-ray excess and cosmic ray rate well enough that it can serve as a reference for both. In the case of the cosmic ray rate, this can be used as a reference for the ideal performance of the instrument. This allows for a precise monitoring of the instrument's performance as suggested already in~\cite{bib:hildebrand1} and the stability of the cosmic ray spectrum allows to derive a quality parameter, for example, on the weather conditions. This parameter is then exclusively obtained from the data itself without any further atmospheric monitoring. For gamma-rays, it allows to obtain a zenith-angle dependent efficiency correction directly from a fit to the data of a stable source as the Crab Nebula instead of requiring a time-consuming detector simulation which usually requires a lot of fine-tuning. It might even be possible to use the obtained zenith-angle dependence to cross-check and fine-tune the Monte Carlo production.

%Thus, all three parameters are related to meaningful physics quantities and changes of th

Once the zenith angle dependence is understood, additional effects can be taken into account as general weather conditions or the Saharan Air Layer as discussed in~\cite{bib:sal}.

% https://www.fact-project.org/dch/db_explorer.php?short=5490

\section*{Acknowledgments} This work is based on previous works by M.~Mahlke. I would like to thank the whole FACT Collaboration allowing me to use our already published data in my own work and all my colleagues who provided valuable input, comments and suggestions.

\newcommand{\arXiv}[1]{{[\href{http://arxiv.org/abs/#1}{arXiv:#1}]}}
\newcommand{\ATel}[1]{{\href{http://www.astronomerstelegram.org/?read=#1}{Astronomer's Telegram~\##1}}}
\newcommand{\doi}[1]{{[\href{http://dx.doi.org/#1}{doi:#1}]}}
\newcommand{\PoS}[3]{{[\href{http://pos.sissa.it/#1/#3/}{PoS(#2)#3}]}}
\newcommand{\DOIarXiv}[2]{{[\href{http://dx.doi.org/#1}{doi:#1}, \href{http://arxiv.org/abs/#2}{arXiv:#2}]}}
\newcommand{\PoSarXiv}[4]{{[\href{http://pos.sissa.it/#1/#3/}{PoS(#2)#3},
\href{http://arxiv.org/abs/#2}{arXiv:#4}]}}

\end{document}